\documentclass[runningheads]{llncs}
\usepackage{graphicx}
\usepackage{amsmath,amssymb} 
\usepackage{subcaption}
\usepackage{tabularx}
\usepackage{tikz}
\usepackage{cite}
\usepackage{marvosym}
\usepackage[utf8x]{inputenc}
\usepackage{algorithm}
\usepackage{algorithmic}
\usepackage{threeparttable}
\usepackage{tablefootnote}
\usepackage[normalem]{ulem}
\usepackage[T1]{fontenc}

\begin{document}
\title{Enhancing Community Vision Screening: AI-Driven Retinal Photography for Early Disease Detection and Patient Trust}

%
%
\author{Anonymized}
\author{Xiaofeng Lei\inst{1} \and
Yih-Chung Tham\inst{2,3} \and
Jocelyn Hui Lin Goh \inst{2} \and
Yangqin Feng\inst{1} \and
Yang Bai\inst{1} \and
Zhi Da Soh\inst{3} \and
Rick Siow Mong Goh \inst{1}\and
Xinxing Xu\inst{1}\textsuperscript{\Letter} \and
Yong Liu\inst{1}\textsuperscript{\Letter} \and
Ching-Yu Cheng \inst{2,3}
}

\authorrunning{X. Lei et al.}
\titlerunning{Enhancing Community Vision Screening}
%
\institute{Institute of High Performance Computing (IHPC), Agency for Science, Technology and Research (A*STAR), Singapore
\newline\email{lei\_xiaofeng@ihpc.a-star.edu.sg}
\and
Singapore Eye Research Institute, Singapore National Eye Centre, Singapore
\and
Department of Ophthalmology, Yong Loo Lin School of Medicine, NUS, Singapore
}

\maketitle              
\begin{abstract}
Community vision screening plays a crucial role in identifying individuals with vision loss and preventing avoidable blindness, particularly in rural communities where access to eye care services is limited. Currently, there is a pressing need for a simple and efficient process to screen and refer individuals with significant eye disease-related vision loss to tertiary eye care centers for further care. An ideal solution should seamlessly and readily integrate with existing workflows, providing comprehensive initial screening results to service providers, thereby enabling precise patient referrals for timely treatment.
This paper introduces the Enhancing Community Vision Screening (ECVS) solution, which addresses the aforementioned concerns with a novel and feasible solution based on simple, non-invasive retinal photography for the detection of pathology-based visual impairment. 
Our study employs four distinct deep learning models: RETinal photo Quality Assessment (RETQA), Pathology Visual Impairment detection (PVI), Eye Disease Diagnosis (EDD) and Visualization of Lesion Regions of the eye (VLR). We conducted experiments on over 10 datasets, totaling more than 80,000 fundus photos collected from various sources. The models integrated into ECVS achieved impressive AUC scores of 0.98 for RETQA, 0.95 for PVI, and 0.90 for EDD, along with a DICE coefficient of 0.48 for VLR. These results underscore the promising capabilities of ECVS as a straightforward and scalable method for community-based vision screening.
\end{abstract}
\section{Introduction}
Visual impairment (VI) is a major public health problem. Globally, there are approximately 2.2 billion people with vision impairment and of which, at least 1 billion were preventable. The top causes of vision impairment include cataract, undercorrected refractive error, age related macular degeneration(AMD), glaucoma, diabetic retinopathy (DR) and myopic macular degeneration (MMD)\cite{citeWHO, tm2012, GBD_2019}.

There is a disproportionate number of ophthalmologists to the rapidly ageing population\cite{Resnikoff588}. Vision screening allows early detection of potential vision threatening eye conditions so that patients can undergo treatment to prevent or slow down the progression of their conditions. However, eye care is poorly integrated within health systems, and the costs of eye care services are not covered by health insurance schemes in many low- and middle-income countries (LMICs)\cite{citeWHO}. The ideal vision screening solution needs be \emph{simple, reliable and cost-efficient}.

Recent studies utilised deep learning in eye disease and systemic disease detection using retinal photography and yielded notable performance\cite{Milea2020-tu-full, tham2021referral, LI2021101971, lei2022localizing}. AI has a direct impact and is transforming eye care\cite{sm2023implementing, Petersson2022}. However, in addition to the predictive performance, the current challenges of AI in eye care also include feasibility tests, and the effectiveness in actual clinical settings as it is difficult to ensure seamless integration of AI tools into existing clinical workflows to improve accessibility to eye care services\cite{ml_translation, google_dr_deploy}. In reality, most eye care services are provided by tertiary hospitals or eye care centers, and thus in rural regions with geographical barriers and scarce health care resources, access to eye care is limited. The ideal vision screening solution needs be \emph{easy in implementation and integration and cater to the large-scale community-based setting}.

To address the above challenges and meet the requirements of large-scale screening, we propose the Enhancing Community Vision Screening (ECVS) solution, which utilizes AI-driven retinal photography. The main contributions are summarized as follows:

\noindent \textbf{1) We propose a novel community vision screening solution with a one-stop workflow tailored for large-scale settings, distinguishing it from typical vision screening approaches.}

\begin{figure}[!h]
\centering
\includegraphics[width=0.9
\textwidth]{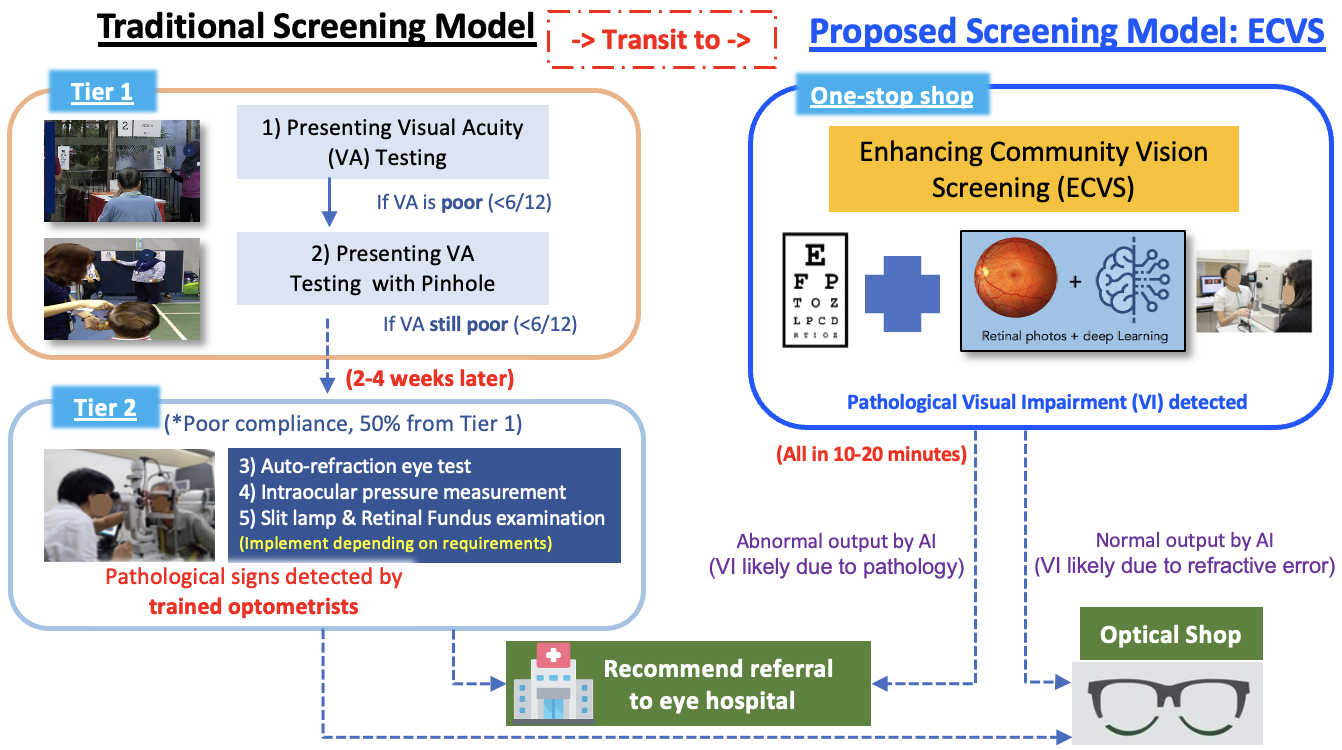}\centering
\caption{Comparison of the current and proposed screening model.} \label{fig1}
\end{figure}

In many countries, traditional vision screening involves a two-tier process. Initially, individuals undergo a visual acuity (VA) test, followed by a pinhole test if their VA is poorer than 6/12. Those with persistently poor VA after the pinhole test are referred to eye professionals for comprehensive screening, including VA and refraction tests, and an eye health examination. Pathology VI cases identified during this process are then referred to eye hospitals as shown in Fig.~\ref{fig1}. However, this model is not sustainable, and not well suited to the large-scale community-based screening. 

To address this challenge, we propose an AI-driven one-stop solution called ECVS for mass screening of pathology VI. ECVS can be deployed at general practitioner (GP) offices, polyclinics, and community centers, including optical shops. Compared to the traditional workflow, ECVS reduces the number of tests from five to two objective tests, decreases total chair time from 40 minutes to just 5 minutes, and reduces referral time from 2-4 weeks to 10-20 minutes in a trial testing.

\noindent \textbf{2) We propose a deep learning pathology VI detection model to facilitate efficient vision screening in communities.}

In the actual clinical setting, less than half of the screening-positive individuals will go to see an eye doctor due to the waiting time. Hence, the follow-up rate is very poor. To overcome this issue, we utilize pathology AI detection model on retinal photos (Pathology VI was defined as presence of eye diseases with best-corrected visual acuity worse than 6/12) which is able to provide a screening report on-site in during the visit, thus reducing manpower and healthcare costs significantly.

\noindent \textbf{3) We propose a screening model featuring a novel visualization module that incorporates a segmentation model and initial diagnosis, thereby improving transparency and fostering trust in screening outcomes.}

A vision screening is an eye test aimed at detecting potential vision problems. Upon discovering an underlying eye condition or visual impairment, concerns often arise regarding its cause and consequences. We provide the results of an initial diagnosis to reassure the patient and assist subsequent specific diagnoses. Additionally, a segmentation model is employed to identify the region of pathology VI which can provide some visual aids to the operator.

\section{Method}
\begin{figure}[!h]
\centering
\includegraphics[width=1.0
\textwidth]{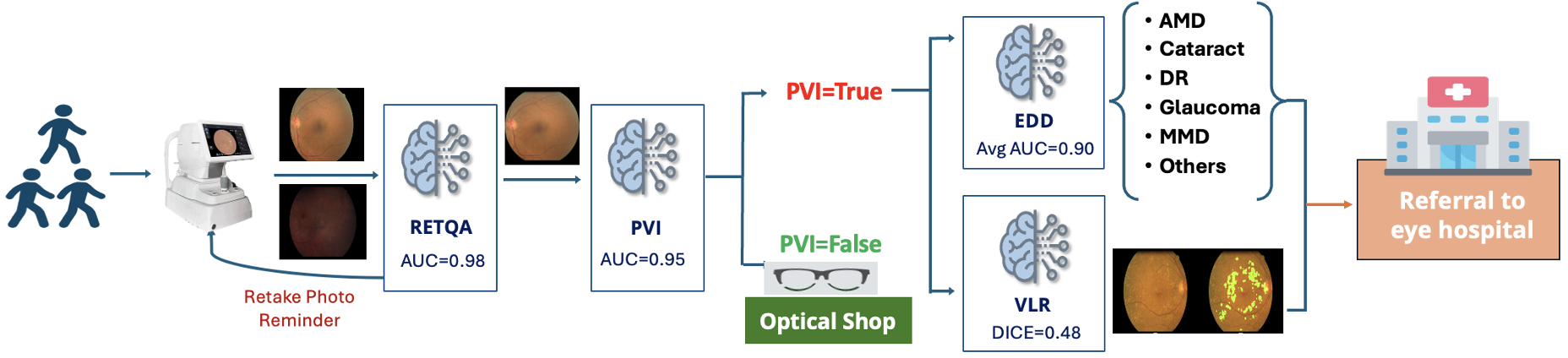}\centering
\caption{The framework of the Enhancing Community Vision Screening (ECVS).} \label{fig2}
\end{figure}

We employed four deep learning models, spanning from the acquisition of retinal photos to the generation of screening results in the proposed ECVS. As shown in Fig.~\ref{fig2}, (1) One RETinal photo Quality Assessment (RETQA) classification model is employed to assess the quality of acquired images. Upon detection of poor quality images, the operator receives a prompt to re-capture the images; (2) The Pathology Visual Impairment (PVI) classification model will give a VI detection with an possibility score. One threshold strategies for vision impairment detection will be tailored to the unique needs and context of community setting; (3) The multi-label eye disease diagnosis (EDD) classification works on the "visual impairment" cases, making the initial diagnosis working like a junior ophthalmologist; (4) The visualization of lesion region (VLR) segmentation model highlights the suspicious region on the "visual impairment" images. Ultimately, the screening results, including the status of PVI and retinal photos, will be generated. Suspected PVI patients will receive an initial diagnosis with lesion region visualization and will then be referred to an eye hospital for further diagnosis.

We employed various backbones, namely ResNet\cite{Resnet}, EffcientNet\cite{efficientnet}, Vision Transformer (ViT)\cite{vit_method}, Swin Transformer\cite{swin_v2_method} and RetFound\cite{Zhou2023-bt}  within the classification model to conduct a comparative performance analysis. For data augmentation, we applied random horizontal flipping, scaling, rotation and gamma correction\cite{data_aug_omia}. U-Net network was adopted in the lesion region segmentation model\cite{unet}. The initial weights of the base networks are loaded from pre-trained models based on ImageNet. 
\subsection{Retinal photo quality assessment (RETQA) classification model}
Retinal image quality assessment is essential in automated analysis systems and it has been well explored\cite{Lin2020-ox}. We compiled a total of 46,320 fundus images which were used for training 3 selected backbones (ResNet50, EfficientNet b7 and Swin V2). The training set comprised 12,543 images from the open-source EyeQ’s training set\cite{huazhu_miccai_qa}, as well as 26,028 images from the UK Biobank (UKBB)\cite{ukbb_dataset} and 7749 images from the Singapore Epidemiology of Eye Diseases (SEED) study\cite{seed_dataset}. For evaluation of the trained model performance, we used 16,249 images from the public EyeQ’s test set\cite{huazhu_miccai_qa}. External validation was conducted using one population study, 6663 post-dilation images from the Beijing Eye Study (BES)\cite{bes_dataset} and one clinical study, 410 pre-dilation images captured by Crystalvue camera from local clinic (Crystalvue QA).
\subsection{Pathology Visual Impairment (PVI) classification model}
In contrast to the VI work in \cite{tham2021referral}, several recent approaches, namely ViT-B, Swin V2, and a ViT-L model pretrained on RetFound, are being explored in the same setting with the aim of achieving better and more reliable performance. Additionally, one more clinical study has been added for external validation. Training involved 15,227 images from the SEED study\cite{seed_dataset}, with an additional 3,803 images for testing. we further validated it using five external independent datasets. We used three population-based studies: 6,239 images from the Beijing Eye study(BES)\cite{bes_dataset}, 6,526 images from Central India Eye and Medical study (CIEMS)\cite{ciems_dataset}, and 2,003 images from Blue Mountains Eye Study (BMES)\cite{bmes_dataset}; and two clinical studies: 968 images from the Chinese University of Hong Kong’s Sight Threatening Diabetic Retinopathy study (CUHK-STDR)\cite{cuhk_dataset}
and 370 pre-dilation images captured by Crystalvue camera from local clinic (Crystalvue VI).
\subsection{Eye disease initial diagnosis (EDD) classification model}
To provide the initial diagnosis for VI cases, we selected the top 5 eye diseases (cataract, AMD, glaucoma, DR, and MMD) causing distance vision impairment or blindness, and included an "others" category as the output in this multi-label classification model. We compiled a total of 11,602 fundus images from the SEED study\cite{seed_dataset}, comprising 6 categories labels used for training 3 selected backbones (ResNet50, ViT-B, and Swin V2). Among these images, there were a total of 8,588 multi-labeled images. Specifically, 800 images were labeled for AMD, 1,797 images for Cataract, 935 images for DR, 2,794 images for Glaucoma, 5,981 images for MMD, and 9,588 images for Others (including other eye diseases and non-PVI images). We randomly split SEED dataset as We randomly split SEED dataset as Train:Val:Test=80$\%$:10$\%$:10$\%$. External validation was conducted using three population study, 457 DR images out of 1744 images (Moderate retinopathy, Severe retinopathy and Proliferative retinopathy category) from MESSIDOR-2\cite{messidor_dataset}, 77 cataract images out of 4881 images from BES\cite{bes_dataset} and 155 glaucoma images out of 489 images (Suspected glaucoma and Glaucoma category) from PALILA\cite{papila_dataset}.
\subsection{Visualization of lesion regions (VLR) segmentation model}
\begin{figure}[!h]
\centering
\includegraphics[width=1.0
\textwidth]{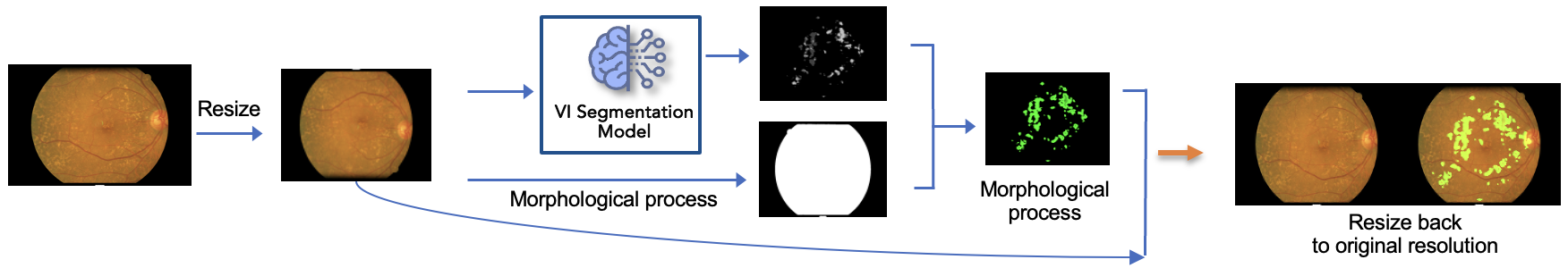}\centering
\caption{The workflow of the lesion region visualization using segmentation model.} \label{fig3}
\end{figure}

The visual explanation of decisions made by classifiers remains an open challenge. Existing visualization methods have limitations, including their reliance on correct categorical labels to identify regions of interest \cite{gradcam, saliency_comparison_2023-full, yiming_heatmap}. Additionally, the saliency maps generated are not consistent, making them rarely used in practical applications. One category-independent method, Hessian-CIAM, has been proposed\cite{yiming_heatmap}, it also faces the same limitation in outputting consistent and reliable region of interest (ROI) heatmaps.
To tackle this issue, we introduce a segmentation model into the visualization process, as illustrated in Fig.~\ref{fig3}. It combines the mask output from the segmentation model with the results of morphological processing of the retinal photo, aiming to refine the output and enhance its interpretability for human understanding in a category-independent manner.
We utilized the 459 annotated images for the various signs and symptoms (Macular hole, drusen, Macular oedema, and so on) from the SEED study\cite{seed_dataset} for training U-NET backbone. We randomly split SEED dataset as Train:Val:Test=80$\%$:10$\%$:10$\%$. External validation was not conducted due to the unavailability of data for calculating evaluation metrics.
\section{Results and Discussion}
\subsection{Model performance}
All models were implemented using the PyTorch framework, and training was conducted utilizing the Adam optimizer. We utilize the Area Under the receiver operating characteristic Curve (AUC) as a metric to evaluate the performance of the classification models, while utilizing the Dice coefficient (DICE) for evaluating the segmentation model.

\noindent \textbf{RETQA model:}
The Swin V2 model achieved an AUC of 0.9889 and 0.9848 on the two internal test sets, with AUCs ranging from 0.9267 to 0.9281 across external test sets. The ResNet50 and EfficientNet b7 network showed a comparable performance on internal and external test set in Table~\ref{tab1}.
\begin{table}[!t]
\centering
\caption{Comparison of RETQA performance across methods}
\begin{threeparttable}
\begin{tabular}{|l|c|c|c|c|}
\hline
Method measured by AUC & ResNet50   & EfficientNet b7 & Swin V2\\
\hline
\multicolumn{4}{|l|}{Primary validation} \\
\hline
SEED+EyeQ+UKBB test &  0.9875 & 0.9881 & 0.9889\\
EyeQ test &  0.9844& 0.9818 & 0.9848 \\
\hline
\multicolumn{4}{|l|}{External testing} \\
\hline
BES &  0.9292 & 0.9049 & 0.9267 \\
Crystalvue QA& 0.9704 & 0.9370 & 0.9281 \\
\hline
\end{tabular}
\end{threeparttable}
\label{tab1}
\end{table}

\noindent \textbf{PVI classification model:}
The Swin V2 model achieved an AUC of 0.9514 on the internal test set, with AUCs ranging from 0.8486 to 0.9593 across external test sets. The ViT-B and RetFound network showed a comparable performance on internal and external test sets in Table 2.
\begin{table}[!t]
\centering
\caption{Comparison of PVI performance across methods}
\begin{threeparttable}
\begin{tabular}{|l|c|c|c|c|}
\hline
Method measured by AUC & ResNet50+Xgboost\tnote{+}  & ViT-B & Swin V2 & RetFound\cite{Zhou2023-bt}\\
\hline
\multicolumn{5}{|l|}{Primary validation} \\
\hline
SEED test &  0.942 & 0.9428 & 0.9514& 0.9497\\
\hline
\multicolumn{5}{|l|}{External testing} \\
\hline
BES &  0.936 & 0.9260 & 0.9194 & 0.9317 \\
CIEMS& 0.928 & 0.9453 & 0.9515 & 0.9593 \\
BMES& 0.890 & 0.9310 & 0.9306 & 0.9307 \\
CUHK-STDR& 0.866 & 0.8820 & 0.8836 & 0.8769 \\
Crystalvue VI& NA\tnote{+} & 0.8630 & 0.8293 & 0.8486 \\
\hline
\end{tabular}
\begin{tablenotes}
    \footnotesize
    \item[+] The results in this column come from \cite{tham2021referral}, no result for Crystalvue VI dataset.
\end{tablenotes}
\end{threeparttable}
\end{table}

\noindent \textbf{Multi-label EDD classification model:}
The Swin V2 model achieved an mean AUC of 0.8964 on the internal test set. The ResNet50 and ViT-B model achieved AUC of 0.8876 and 0.8511 respectively. The Swin V2 model also achieved an mean AUC of 0.8634 on the external test sets as shown in Table 3.

\begin{figure}[!h]
\centering
\includegraphics[width=1.0
\textwidth]{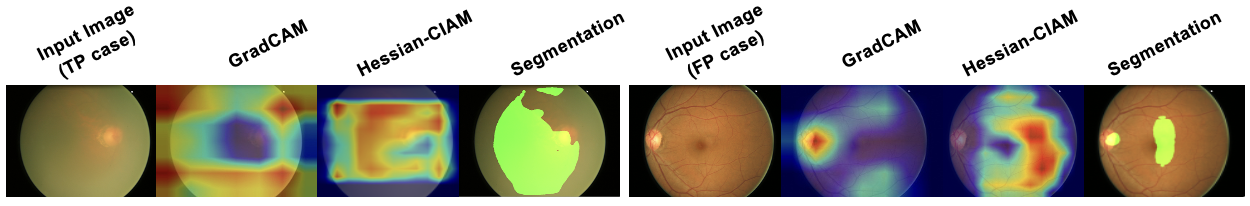}\centering
\caption{Saliency maps for PVI visualization using different methods}\label{fig4}
\end{figure}

\begin{table}[!t]
\centering
\caption{Comparison of Multi-label EDD model performance across methods}
\begin{threeparttable}
\begin{tabular}{|l|c|c|c|c|c|c|}
\hline
 & \multicolumn{6}{|c|}{SEED Dataset} \\
\hline
Method measured by AUC & AMD & Cataract & DR & Glaucoma  & MMD & Others\\
\hline
\multicolumn{7}{|l|}{Primary validation} \\
\hline
ResNet50 &  0.9006 & 0.8936 & 0.9518 & 0.7575 & 0.9438 & 0.8785\\
ViT-B &  0.8145 & 0.8881 & 0.9020 & 0.7027 & 0.9340 & 0.8656\\
Swin V2 &  0.9220 & 0.9001 & 0.9714 & 0.7562 & 0.9487 & 0.8796\\
\hline
External testing & & & & & &\\
\hline
Swin V2 & NA\tnote{+}  & 0.8241 & 0.9234 & 0.8428 & NA\tnote{+} & NA\tnote{+}\\
\hline
\end{tabular}
\begin{tablenotes}
    \footnotesize
    \item[+] Only Cataract, DR, and Glaucoma photos available as inputs.
\end{tablenotes}
\end{threeparttable}
\end{table}
\noindent \textbf{VLR segmentation model:}
The U-NET model achieved a DICE score of 0.482 on the internal test set. The performances of the method are reported in Table 4. Examples of saliency maps are presented in Fig.~\ref{fig4}.
\begin{table}[!t]
\centering
\caption{VLR segmentation model performance using U-NET.}
\begin{threeparttable}
\begin{tabular}{|l|c|c|c|c|}
\hline
Method measured by DICE score& U-NET \\
\hline
\multicolumn{2}{|l|}{Primary validation} \\
\hline
SEED test &  0.482 \\
\hline
\end{tabular}
\end{threeparttable}
\end{table}
\subsection{Discussion}
From the results, we could have several observations:

1) The RETQA models have AUCs of over 0.9 on both internal and external datasets, including post-dilation and pre-dilation retinal photos. This demonstrates that the RETQA model accurately determines gradability and makes reliable decisions regarding image quality control.

2) The PVI models achieved an AUC of over 0.94 on the internal dataset and an AUC of over 0.82 on the external datasets. With an appropriate threshold, it can ensure that the PVI output is tailored to the context of a community setting with high sensitivity or specificity. Our next plan involves introducing the concept of uncertainty, which could further enhance the reliability of PVI detection in future deployments\cite{Wang2023-mt-full}.

3) The multi-label eye disease classification model provides initial diagnoses akin to those made by a junior ophthalmologist working in the community. The Swin V2 model achieved a mean AUC of 0.896 on the internal dataset and a mean AUC of 0.863 on the external datasets. Given that these diagnoses are provided during the screening stage, the results are considered acceptable.

4) The segmentation model introduced a novel approach to visualizing the eye lesion region. In comparison to other gradient-based methods, the ROI heatmap generated by the segmentation model is more reasonable and aligns closely with human perception in both TP and FP cases, as illustrated in Fig.~\ref{fig4}. We achieved a DICE score of 0.482 required trained on 459 annotated images, with potential for further improvement through additional training data.

\textbf{Computational Efficiency}: The ECVS comprises four deep learning models, which are executed on the CPU during community deployment for cost-efficiency. For instance, when Swin V2 and U-Net networks are run with a batch size of 1 and no additional workers, CPU RAM usage remains below 4GB per model. Furthermore, the inference time is faster than 1 image per second on a standard laptop PC equipped with an Intel Core i5 processor and 16GB DDR memory. Compared to traditional vision screening processes, the computational overhead of the ECVS is manageable and feasible.

Our study has several limitations. Firstly, while our training dataset was diverse and we conducted external validation, there is room for improvement in dataset diversity, innovation, and depth of analysis. Further validation across additional demographics is needed to enhance the generalizability of these four distinct deep learning models.
Secondly, the Visualization of Lesion Region (VLR) segmentation differs from gradient-based saliency map approaches and shows promise. However, external validation was hindered by data unavailability. Future studies should explore larger datasets to enhance and validate this method.
Lastly, our solution involves a visual acuity test conducted by trained healthcare professionals, which may compromise patient trust and screening reliability. Future research should focus on developing robust deep learning models to replace visual acuity tests, thereby enhancing screening efficiency and intelligence.

\section{Conclusions}
In this paper, we present an innovative solution, the Enhancing Community Vision Screening (ECVS), addressing key challenges encountered in existing screening programs. These challenges include prolonged screening cycles, limited human resources, and inadequate coverage of the general public. Our proposed solution is characterized by its simplicity, efficiency, and suitability for large-scale community screening settings.

We employed a sequential integration of four deep learning models, encompassing the entire process from photo acquisition to the generation of screening results. These models encompass retinal image quality assessment (RETQA), pathology visual impairment (PVI) detection, Eye Disease Diagnosis (EDD) and Visualization of Lesion Regions of the eye (VLR). The results obtained from each model within our proposed solution demonstrate significant promise, indicating a potential avenue for future advancements in community vision screening leveraging retinal photos and AI technologies. While our proposal offers evident advantages over existing approaches in terms of cost and feasibility, further evaluation of AI's performance in real-world settings is warranted. This evaluation should encompass factors such as time efficiency, patient volume considerations, and patient acceptance during real-world implementation.

\subsection*{Acknowledgements. 
{\normalfont This work was supported by the Agency for Science, Technology and Research (A*STAR) under its AME Programmatic Funds (Grant Number : A20H4b0141), and its RIE2020 Health and Biomedical Sciences (HBMS) Industry Alignment Fund Pre-Positioning (IAF-PP, Grant Number : H20c6a0031).}}

\bibliographystyle{ieeetr}
\bibliography{Paper-0033}
\end{document}